# Ultrafast optical Kerr Gate at 1 GHz repetition rate by focusing on BBS glass

*Amr Farrag*[1], Assegid M. Flatae[1], Lenorah M. Stott[2], Alessandro Jagatti[3,5], Andrea Lapini[3,4], Doris Möncke[2], Mario Agio[1,3,5]*

1. Laboratory of Nano-Optics, University of Siegen, 57072 Siegen, Germany
2. Inamori School of Engineering at the New York State College of Ceramics, Alfred University, Alfred, NY, 14802, USA
3. European Laboratory for Non-Linear Spectroscopy, 50019 Sesto Fiorentino, FI, Italy
4. Dipartimento di Scienze Chimiche, della Vita e della Sostenibilità Ambientale, Università di Parma, 43124 Parma, PR, Italy
5. Istituto Nazionale di Ottica (INO), Consiglio Nazionale delle Ricerche (CNR), 50019 Sesto Fiorentino (FI), Italy

*E-mail: amr.farrag@uni-siegen.de; mario.agio@uni-siegen.de

Keywords: ultrafast phenomena, optical non-linearity, third-order nonlinear process, BBS glass

**Abstract:** Efficient detection of ultrafast phenomena is central to modern optical sciences, driving advances in quantum science and technology, physical chemistry, and nanophotonics. When processes occur on sub-picosecond timescales, time-resolved methods such as transient absorption, up-conversion, and optical Kerr gating (OKG) can be utilized to probe these dynamics, though they are typically applied to ensembles rather than single emitters. The OKG technique offers high detection efficiency over a broad spectral range, making it particularly promising for ultrafast single-photon detection at high-repetition-rates. Here, we present an ultrafast scheme based on the third-order nonlinearity of bismuth borosilicate (BBS) glass, operating at a 1 GHz repetition rate with less than 1 nJ pulse energy under focusing, and achieving a time resolution down to 175 fs. BBS glass was selected for its high nonlinear coefficient, which enhances detection efficiency, sub-ps response time, and its compatibility with standard microscopy platforms.

1.  Introduction

Measuring ultrafast phenomena lies at the forefront of modern optical science, unlocking new capabilities in fields such as quantum information processing, high-sensitivity detection, and remote sensing [1-4]. Beyond these technological applications, it also serves as a powerful analytical tool for life sciences, enabling investigations at the single-molecule level [5]. However, certain emission processes occur on timescales too short to be resolved with conventional methods, for instance, excited-state dynamics in porphyrins [6] and β-carotene [7]. At the same time, ongoing advances in hybrid quantum systems, where quantum emitters are coupled to nanophotonic structures, show that sub-picosecond emission dynamics are increasingly within experimental reach [8-11]. These developments require advanced ultrafast techniques capable of probing photodynamics at the single-emitter level. To this end, time-resolved ultrafast spectroscopy methods have been developed, such as transient absorption, up-conversion, and optical gating. Yet, to date, these techniques have primarily been applied to molecular ensembles or attenuated laser sources rather than to individual quantum emitters [12-16].

Conventional time-resolved spectroscopy for determining fluorescence lifetimes requires measuring photon emission dynamics following pulsed excitation. Time-Correlated Single Photon Counting (TCSPC) reconstructs decay profiles by accumulating single-photon events over many cycles [17]. The method relies on single-photon-sensitive detectors such as single-photon avalanche diodes (SPADs) or superconducting nanowire single-photon detectors (SNSPDs), whose dead times limit count rates to tens of MHz. While modern TCSPC electronics reach input rates above 1 GHz, the overall temporal resolution, typically a few ps, is constrained by detector jitter and instrument response [18]. Although TCSPC enables picosecond-scale measurements, sub-picosecond processes are still inaccessible. Streak camera–based approaches achieve resolutions down to a few picoseconds and even 100 fs in single-shot mode [19-21], but their bulkiness, cost, and instability at high repetition rates hinder practical use for compact, high-speed single-photon characterization.

Recent advances in stable ultrafast lasers, such as sub-100 fs Ti:sapphire systems, and associated optoelectronic tools have enabled time resolutions comparable to the excitation pulse width through nonlinear optical sampling. Among these, fluorescence up-conversion, based on second-order nonlinear processes in crystals such as KDP, $LiNbO_3$, or BBO, provides the temporal precision needed to study ultrafast events across the UV to near-infrared range [14, 22–28]. This technique achieves resolutions down to 33 fs, making it one of the most precise time-resolved methods. However, its dependence on factors such as phase matching, group velocity mismatch, and crystal parameters limits sensitivity and efficiency for single-photon detection, despite recent advances in single-photon frequency upconversion [4].

An alternative and promising nonlinear sampling method, the optical Kerr gate (OKG), sometimes also referred to as optical Kerr shutter (OKS) offers comparable sub-picosecond resolution without requiring phase matching. It is based on the third-order nonlinear Kerr effect and relies on induced birefringence in materials such as YAG, GGG, BGO, or fused silica. Time resolutions of 80–100 fs and Kerr efficiencies above 90% have been reported [12, 30]. The OKG setup allows temporal gating of fluorescence through polarization rotation, providing ultrafast single-photon detection with high efficiency and broad spectral flexibility. In general, the OKG technique combines high detection efficiency across a broad spectral range. These attributes make it particularly promising for ultrafast single-photon detection, where the emitted signal from a single quantum emitter is inherently weak. Moreover, maintaining a high repetition rate is equally critical to ensure sufficient photon statistics and signal-to-noise ratio in time-resolved measurements.

In this paper, we demonstrate an ultrafast detection scheme, namely an OKG, with an instrument response function of about 175 fs. In addition, the OKG operates at 1 GHz with less than 1 nJ pulse energy. This rate is orders of magnitude larger and the pulse energy is orders of magnitude smaller than literature results. This is possible based on light focusing [30] and the ultrafast strong third-order nonlinearity of bismuth-borosilicate (BBS) glass [31,32].

## 2. Experimental approach and discussions

In the following, we discuss the sample fabrication and the experimental approach. The BBS used for our optical Kerr gate experiments is a ternary system composed primarily of $Bi_2O_3$-$B_2O_3$-$SiO_2$ (BBS). Two BBS glass samples were prepared with the following composition: $Bi_2O_3$ (60 mol%), $B_2O_3$ (20 mol%), and $SiO_2$ (20 mol%). Here, the mol% denotes the nominal mole fraction relative to the total moles of all components. This composition was adopted from a previous study on the third-order nonlinear optical properties of BBS glass [31].

Unlike Ref. [31], $CeO_2$ was excluded, and the glass formation conditions were deliberately modified to minimize coloration. The raw ingredients were melted in an $Al_2O_3$ crucible at 950 °C for 30 min, followed by annealing at 400 °C for 1 h and slow cooling to room temperature. The resulting glass samples were polished to a thickness of 3 mm. Figure 1a shows an image of one of the BBS glass specimens mounted in a holder. Despite external polishing, Fig. 1a reveals that the glass is not internally uniformly cooled, resulting in some thickness (density fluctuation) variations across the sample. However, these variations have a negligible effect on focused light transmission. Transmission measurements using a standard commercial spectrophotometer (PerkinElmer) (Fig. 1b) confirm transparency in the visible and near-infrared regions.

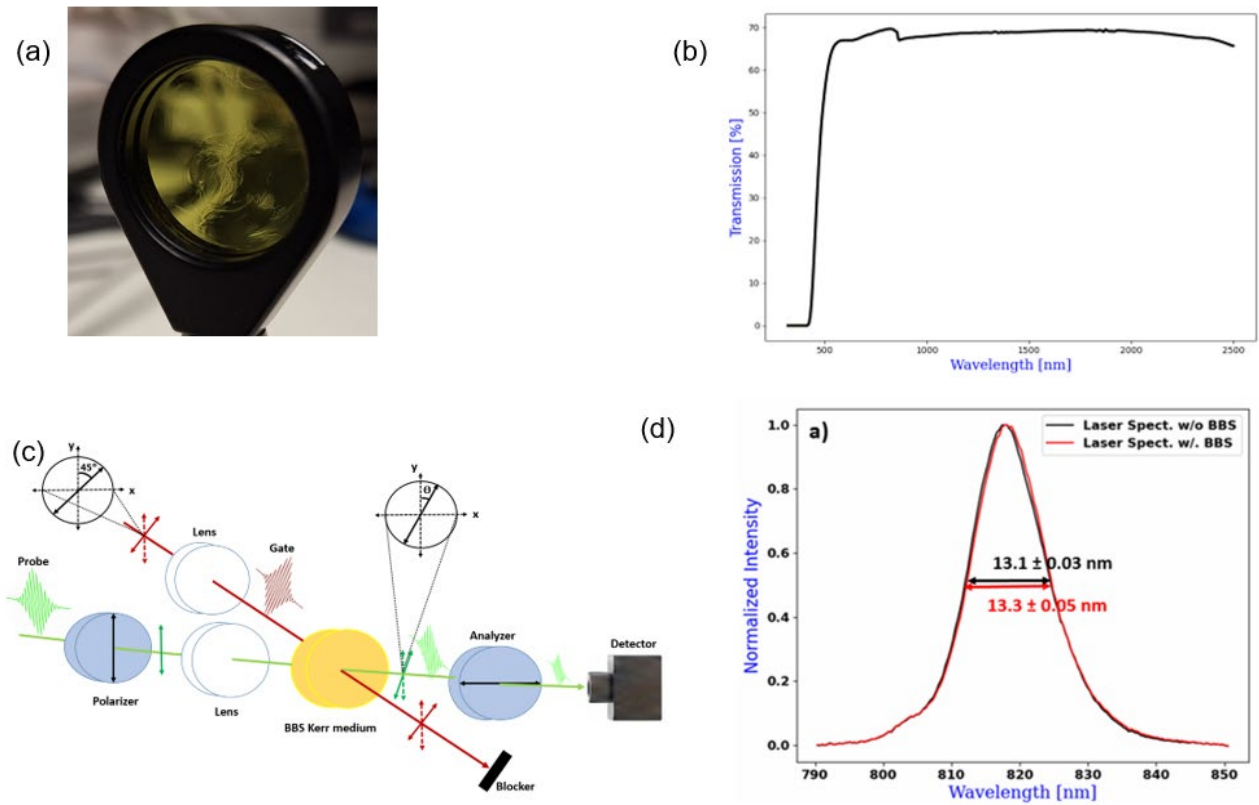

Figure 1. (a) The 3 mm thick BBS glass specimen. (b) Transmission spectrum of BBS glass (spectral artefact at 830 nm is due to the spectrophotometer. (c) Scheme of the experimental setup for Kerr gate experiment. (d) Measured laser spectrum with and without BBS.

For the optical measurements, we employed a tunable ultrafast laser system (Taccor Tune 10, Laser Quantum GmbH) with a tuning range of 730–880 nm, pulse duration < 80 fs, average power of 2.1 W at 800 nm, and a repetition rate of 1 GHz. The experimental setup is schematically shown in Fig. 1c. For simplicity, a degenerate configuration was used for the probe and gate beams. The setup illustrates the optical layout of the technique. The probe beam passes through a shutter consisting of a nonlinear medium placed between two crossed polarizers. Due to the 1 GHz repetition rate, the gate beam's peak pulse energy is relatively low (less than 1 nJ) – a constraint imposed by the Taccor laser's output average power when compared to the amplified pulses used in standard OKG experiments. To enhance the gate efficiency despite this limitation, the probe beam and the pump beams are tightly focused down to 20 µm, and 30 µm, respectively [30]. In the absence of the gate pulse, the medium remains isotropic (i.e., no birefringence), and the probe beam is blocked by the second polarizer. When the gate pulse arrives, the optical Kerr effect induces a transient birefringence in the medium, rotating the probe beam's polarization. Consequently, a portion of the probe light is transmitted through the second polarizer and detected by a photodetector. The detected signal corresponds only to the part of the probe pulse temporally overlapping with the gate pulse. Unlike up-conversion techniques, this detection scheme leaves the probe light frequency unchanged.

Before performing the Kerr-gate measurements, it is important to evaluate the effect of self-phase modulation (SPM). SPM, a self-focusing phenomenon, is responsible for the spectral broadening of high-intensity laser beams. The spectrum of the Taccor fs laser was first measured at 820 nm without the BBS glass. The BBS glass was then inserted into the optical path, and the laser spectrum was measured again for comparison. For the SPM measurements, the gate beam—having a peak pulse intensity of 1.33 GW/cm² and a beam diameter of 30 µm was used. To ensure reliable data and account for fluctuations of the fs laser, each set of spectra was measured 10 times, both with and without the BBS glass. The spectra were then averaged for comparison. As shown in Fig. 1d, the presence of the BBS glass produces almost no spectral broadening.

Following this, degenerate optical Kerr gate experiments were performed using both counter-propagating and co-propagating schemes. To avoid detector saturation and minimize noise, the pump and probe beams were preferentially aligned in the counter-propagating configuration.

The nonlinear optical properties of BBS glass, investigated using a degenerate optical Kerr gate, were previously studied by Lin et al. [32] with a 30-fs laser at 800 nm and a repetition rate of 1 kHz. The nonlinear refractive index was estimated to be $n_2$ = 1.6×10$^{-14}$ cm²/W. Furthermore, the nonlinear response time of BBS glass was measured to be below 90 fs, significantly shorter than that of other third-order ($\chi^3$) materials such as $CS_2$. In another study, Ref. [31], BBS glass was examined using degenerate four-wave mixing (DFWM) with a 200-fs laser at 810 nm and a 200 kHz repetition rate. The measured ultrafast response time was < 200 fs, in excellent agreement with earlier results from the same research group [33]. They also demonstrated an optical Kerr shutter [34] employing BBS glass as the Kerr medium for optical switching using a 1.5 THz gate beam, achieving a response time of < 150 fs.

Additionally, Tan et al. [35] utilized BBS glass as the Kerr medium in a nondegenerate optical Kerr gate for gating spectra from a chirped supercontinuum generated in a sapphire plate. The ultrafast response time of BBS glass in their study, reported as < 85 fs, further confirms its superior temporal resolution compared to materials like $CS_2$.

Although the OKG fundamentally arises from photoinduced birefringence in the Kerr medium, degenerate optical Kerr gate experiments may also exhibit self-diffraction of the pump (gate) and probe beams. This self-diffraction results from the formation of a laser-induced transient grating (LITG), which originates from the third-order nonlinear susceptibility. The effect is strongest when the pump and probe beams share the same polarization. It occurs due to the superposition of two coherent pulses of equal wavelength, producing a spatially modulated energy density pattern in the medium. It becomes significant when the laser pulse duration is comparable to the response time of the nonlinear (Kerr) medium, enabling energy exchange between the two beams. Importantly, the temporal behavior of the self-diffraction arising from LITG depends solely on the correlation time of the interfering laser pulses and does not reflect the intrinsic nonlinear response time of the medium. Hence, it is therefore crucial to distinguish between photoinduced birefringence and self-diffraction to accurately determine the time response of the nonlinear medium. Yan et al. [36] investigated the dependence of Kerr-rotated signal intensity (after the analyzer) on pump power and polarization angle in BBS glass. Due to the ultrafast electronic response associated with electron cloud distortion, the time response of BBS glass can be shorter than, and effectively limited by, the laser pulse width itself, typically on the femtosecond timescale. Consequently, it is practically impossible to completely separate the OKG signal from self-diffraction under these conditions.

The polarization angle between the pump and probe beams was adjusted to 45°, achieved by placing a half-wave plate in the gate beam path. The OKG signal was recorded by measuring the Kerr-rotated probe intensity after the second polarizer (analyzer) as a function of the delay Δτ, introduced by varying the optical path length of the pump beam. The pump beam was delayed by the linear translation stage over a distance, corresponding to 1.44×10$^{-12}$ s (~1.44 ps) with a maximum temporal resolution of 0.66 fs. The resulting time trace of the OKG signal is shown in Fig. 2a, where the x-axis represents the time delay and the y-axis corresponds to the number of counts per second, integrated over a 1 s acquisition period.

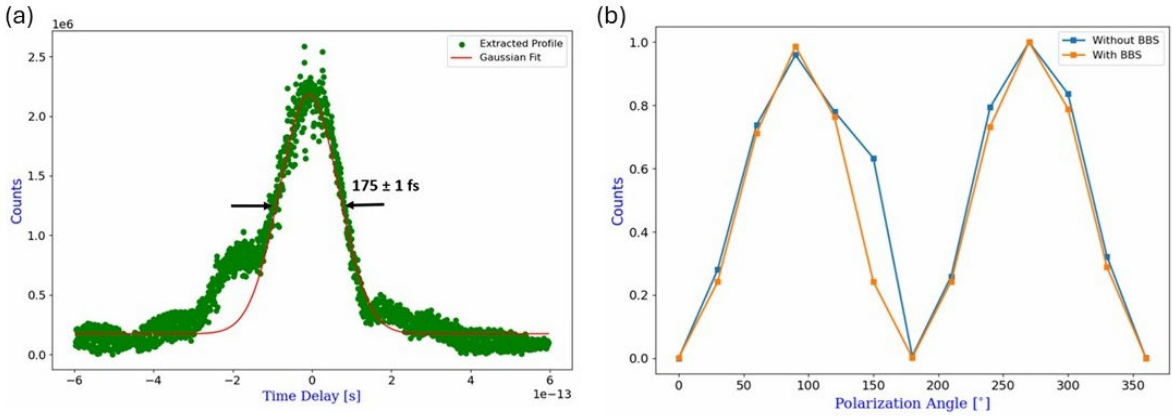

Figure 2. (a) Temporal profile of OKG signal of the time trace, fitted with a Gaussian function, whose FWHM is 175 ± 1 fs. (b) Depolarization measurement of the probe beam (at 760 nm), with/without BBS glass.

The temporal profile shown in Fig. 2a is the envelope function of the OKG signal extracted by Fourier analysis, where the fast oscillatory part of it can be eliminated. The envelope represents the convolution of a cross-correlation between the gate and probe beams, and the nonlinear response time of the BBS glass. This implies that the nonlinear response time of the BBS glass < 175 fs. A precise determination of the actual nonlinear response time would require measurement of the cross-correlation between the pump and probe beams, and deconvolving the temporal profile of the OKG signal via the cross-correlation value. The OKG efficiency of our setup was found to be ~ 3 %, evaluated by a separate experiment (not presented here) in which the probe beam was a CW laser at 656 nm, and the gate beam was the fs laser beam at 820 nm, used for obtaining the above results. We observed a shift in the angle corresponding to the maximum transmitted Kerr signal, which we attribute to ellipticity introduced within the optical setup. Another possible cause considered was that the BBS glass might induce a degree of depolarization, thereby altering the phase of the experimental time traces and their temporal profiles. To investigate this, a half-wave plate was inserted into the probe beam path (at 760 nm). Initially, the BBS glass specimen was removed, and the half-wave plate was rotated through angles ranging from 0° to 360° in 30° increments. The probe beam intensity was recorded using a photodetector, with five measurements taken for each polarization angle. Subsequently, the BBS glass specimen was reinserted into the optical path, and the half-wave plate was placed after the glass, before the analyzer. The same set of measurements was repeated under identical conditions. A comparison of the probe beam intensities with and without the BBS glass is shown in Fig. 2b. As clearly illustrated, the presence of the BBS glass does not introduce any measurable depolarization or phase offset in the probe beam signal.

For applications requiring the detection of extremely weak signals, such as ultrafast single-photon sources, detailed knowledge of the nonlinear material's physical properties is indispensable as the efficiency is critical. One key parameter is the two-photon absorption (TPA) of BBS glass. Linear absorption occurs when the excitation light frequency is resonant with real electronic states. In contrast, within a transparent spectral region far from resonance, two photons can be simultaneously absorbed if the photon density is sufficiently high, a condition typically satisfied within the focal volume of an intense laser beam. When considering nonlinear optical absorption processes such as TPA, the imaginary part of the third-order susceptibility must be considered. Hasegawa et al. [37] measured the nonlinear two-photon absorption coefficient (β) of a bismuth-containing glass (65.5 mol% $Bi_2O_3$) using the Z-scan technique and reported β=0.8 cm/GW at 769 nm. Similarly, Zhang et al. [38] employed the same technique to study a ternary $Bi_2O_3$-rich glass system, $Bi_2O_3$–$B_2O_3$–$WO_3$ (BBW), and reported β values ranging from 0.254 cm/GW to 0.345 cm/GW for a series of samples with varying $WO_3$ content. The higher β value corresponded to the composition with the greatest $WO_3$ concentration, while the lower value was observed for the sample with the least $WO_3$ content.

There are two main reasons why two-photon absorption (TPA) must be carefully considered. First, at high laser intensities, nonlinear absorption reduces the effective light intensity available for inducing refractive index changes, thereby diminishing the efficiency of nonlinear optical processes. Second, nonlinear absorption mechanisms, such as TPA in addition to linear absorption, can become critical when using high-repetition-rate lasers. Even small absorption coefficients can lead to significant heat accumulation within the medium, resulting in the formation of a thermal lens. This thermally induced lens acts as a negative (diverging) lens, opposing or even counteracting both the Kerr-effect-induced self-focusing and the optical focusing of the beam itself. Such effects impose limitations on intensity-dependent nonlinear optical phenomena, including optical Kerr gates, which are commonly used in ultrafast optical switching applications. Considering that the femtosecond laser used in this work (Taccor Tune, 1 GHz repetition rate) operates at more than an order of magnitude higher repetition rate than most commercial solid-state lasers, it is therefore essential to probe TPA in BBS glass.

The presence of TPA in a nonlinear medium is typically investigated by measuring it directly at the wavelength of interest, often alongside the nonlinear refractive index ($n_2$) using techniques such as Z-scan. In our case, direct measurement of $n_2$ or the TPA coefficient ($\beta$) was not feasible for two reasons: (1) implementing a Z-scan setup would require additional optical components and space, and (2) the technique demands higher laser intensities than those provided by the Taccor fs laser.

As an alternative, the possible existence of TPA can be inferred by estimating the optical bandgap of the Kerr medium, in this case, BBS glass, and comparing it with twice the photon energy of the excitation light. If the bandgap corresponds to this value, two-photon absorption is likely to occur. The optical bandgap can be determined using the Tauc plot method [39, 40], which analyzes the linear absorption (or transmission) spectrum of amorphous materials. In this approach, the photon energy is plotted against $(\alpha h\nu)^n$, where n=2 for optically allowed direct bandgap, while n=1/2 for indirect bandgaps, as is appropriate for our systems. The bandgap is obtained from the intercept of the linear portion of the curve with the energy axis. This method has been successfully applied to various glass systems, including BBS glass with different $Bi_2O_3$ contents, as demonstrated by Gao et al. [41]. See Fig. 3, for our Tauc plot curve, along with its linear fitting.

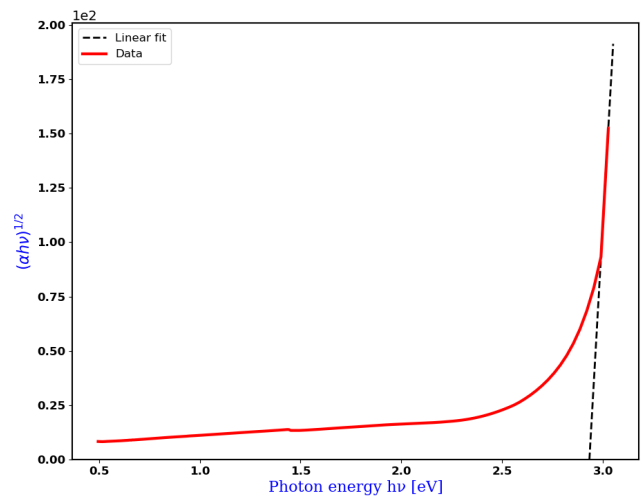

Figure 3. Tauc plot of BBS glass sample. Linear fitting of the Tauc plot of the BBS glass sample that contains 60 % mol. $Bi_2O_3$. This linear fit interception point with x-axis shows that our BBS glass has $E_g$ = 2.93 eV.

For the OKG measurements, the pump (gate) and probe beams were at a central wavelength of 820 nm with a spectral bandwidth of 15 nm. This corresponds to an operational wavelength range of approximately 812.5–827.5 nm. For TPA to occur, two photons within this range must be absorbed simultaneously, equivalent to a wavelength range of 406.25-413.75 nm (half the excitation wavelength). The optical bandgap of our BBS glass specimen determined from Tauc plot (see Fig. 3) is $E_g$=2.93 eV, corresponding to a wavelength of approximately 422.9 nm. Since this bandgap wavelength lies outside the two-photon range (406.25–413.75 nm), it is unlikely that TPA occurs under our experimental conditions. However, it is not completely excluded that the band gap energy be larger than 2.93 eV, because the presence of impurities in BBS appears as a lower band gap [42].

## 3. Conclusion

In summary, we have implemented an ultrafast detection scheme based on the third-order nonlinearity of bismuth borosilicate (BBS) glass, operating at a 1 GHz repetition rate under focusing with less than 1 nJ pulse energy, and achieving an instrument response function with a time resolution of 175 fs. BBS glass was chosen for its large nonlinear coefficient, which enhances detection efficiency and it is necessary for operating at high repetition rates, its sub-ps response time, and its compatibility with standard microscopy platforms. This approach provides a robust platform for probing ultrafast photodynamics at the single-emitter level and holds promise for applications in ultrafast single-photon detection, quantum information processing, high-speed optical switching, and time-resolved spectroscopy in nanophotonic and biological systems.

**Research funding:** This activity has been partially supported by the University of Siegen and the German Research Foundation (DFG) (INST 221/118-1 FUGG, 410405168).

**Author contribution:** All authors have accepted responsibility for the entire content of this manuscript and consented to its submission to the journal, reviewed all the results and approved the final version of the manuscript

**Conflict of interest:** Authors state no conflict of interest.

**Data availability statement**: The datasets generated and analysed during the current study are available from the corresponding author upon reasonable request.

**Acknowledgments:** The authors would like to acknowledge Thomas Lenzer, Kawon Oum, and Giancarlo Soavi for helpful discussion and Mauro Pucci for BBS glass polishing.

## References

1. Y. Arakawa, and M.J. Holmes, "Progress in quantum-dot single photon sources for quantum information technologies: A broad spectrum overview," Applied Physics Reviews 7(2), (2020).

2. V. Shcheslavskiy, P. Morozov, A. Divochiy, Yu. Vakhtomin, K. Smirnov, and W. Becker, "Ultrafast time measurements by time-correlated single photon counting coupled with superconducting single photon detector," Review of Scientific Instruments 87(5), (2016).

3. S. Chan, A. Halimi, F. Zhu, I. Gyongy, R.K. Henderson, R. Bowman, S. McLaughlin, G.S. Buller, and J. Leach, "Long-range depth imaging using a single-photon detector array and non-local data fusion," Sci Rep 9(1), (2019).

4. L. Ma, O. Slattery, and X. Tang, "Single photon frequency up-conversion and its applications," Physics Reports 521(2), 69–94 (2012).

5. E.M.H.P. van Dijk, J. Hernando, J.-J. García-López, M. Crego-Calama, D.N. Reinhoudt, L. Kuipers, M.F. García-Parajó, and N.F. van Hulst, "Single-Molecule Pump-Probe Detection Resolves Ultrafast Pathways in Individual and Coupled Quantum Systems," Phys. Rev. Lett. 94(7), (2005).

6. Y. Venkatesh, M. Venkatesan, B. Ramakrishna, and P.R. Bangal, "Ultrafast Time-Resolved Emission and Absorption Spectra of meso-Pyridyl Porphyrins upon Soret Band Excitation Studied by Fluorescence Up-Conversion and Transient Absorption Spectroscopy," J. Phys. Chem. B 120(35), 9410–9421 (2016).

7. H. Kandori, H. Sasabe, and M. Mimuro, "Direct Determination of a Lifetime of the S2 State of beta-Carotene by Femtosecond Time-Resolved Fluorescence Spectroscopy," J. Am. Chem. Soc. 116(6), 2671–2672 (1994).

8. X.-W. Chen, M. Agio, and V. Sandoghdar, "Metallodielectric Hybrid Antennas for Ultrastrong Enhancement of Spontaneous Emission," Phys. Rev. Lett. 108(23), (2012).

9. T.B. Hoang, G.M. Akselrod, and M.H. Mikkelsen, "Ultrafast Room-Temperature Single Photon Emission from Quantum Dots Coupled to Plasmonic Nanocavities," Nano Lett. 16(1), 270–275 (2015).

10. A.M. Flatae, F. Tantussi, G.C. Messina, F. De Angelis, and M. Agio, "Plasmon-Assisted Suppression of Surface Trap States and Enhanced Band-Edge Emission in a Bare CdTe Quantum Dot," J. Phys. Chem. Lett. 10(11), 2874–2878 (2019).

11. S.I. Bogdanov, O.A. Makarova, X. Xu, Z.O. Martin, A.S. Lagutchev, M. Olinde, D. Shah, S.N. Chowdhury, A.R. Gabidullin, I.A. Ryzhikov, I.A. Rodionov, A.V. Kildishev, S.I. Bozhevolnyi, A. Boltasseva, V.M. Shalaev, and J.B. Khurgin, "Ultrafast quantum photonics enabled by coupling plasmonic nanocavities to strongly radiative antennas," Optica 7(5), 463 (2020).

12. B. Schmidt, S. Laimgruber, W. Zinth, and P. Gilch, "A broadband Kerr shutter for femtosecond fluorescence spectroscopy," Appl. Phys. B 76(8), 809–814 (2003).

13. S. Kinoshita, H. Ozawa, Y. Kanematsu, I. Tanaka, N. Sugimoto, and S. Fujiwara, "Efficient optical Kerr shutter for femtosecond time-resolved luminescence spectroscopy," Review of Scientific Instruments 71(9), 3317–3322 (2000).

14. C.H. Kim, and T. Joo, "Ultrafast time-resolved fluorescence by two photon absorption excitation," Opt. Express 16(25), 20742 (2008).

15. O. Kuzucu, F.N.C. Wong, S. Kurimura, and S. Tovstonog, "Time-resolved single-photon detection by femtosecond upconversion," Opt. Lett. 33(19), 2257 (2008).

16. C. Cimpean, V. Groenewegen, V. Kuntermann, A. Sommer, and C. Kryschi, "Ultrafast exciton relaxation dynamics in silicon quantum dots," Laser & Photonics Reviews 3(1–2), 138–145 (2009).

17. W. Becker, *The Bh TCSPC Handbook*, 8th ed. Becker & Hickl GmbH, 2019.

18. B. Korzh, et al., "Demonstration of sub-3 ps temporal resolution with a superconducting nanowire single-photon detector," Nat. Photonics 14(4), 250–255 (2020).

19. J. Wiersig, C. Gies, F. Jahnke, M. Aßmann, T. Berstermann, M. Bayer, C. Kistner, S. Reitzenstein, C. Schneider, S. Höfling, A. Forchel, C. Kruse, J. Kalden, and D. Hommel, "Direct observation of correlations between individual photon emission events of a microcavity laser," Nature 460(7252), 245–249 (2009).

20. M. Aßmann, F. Veit, M. Bayer, C. Gies, F. Jahnke, S. Reitzenstein, S. Höfling, L. Worschech, and A. Forchel, "Ultrafast tracking of second-order photon correlations in the emission of quantum-dot microresonator lasers," Phys. Rev. B 81(16), (2010).

21. M. Aßmann, F. Veit, J.-S. Tempel, T. Berstermann, H. Stolz, M. van der Poel, J.M. Hvam, and M. Bayer, "Measuring the dynamics of second-order photon correlation functions inside a pulse with picosecond time resolution," Opt. Express 18(19), 20229 (2010).

22. M. Sajadi, M. Quick, and N.P. Ernsting, "Femtosecond broadband fluorescence spectroscopy by down- and up-conversion in β-barium borate crystals," Applied Physics Letters 103(17), (2013).

23. L. Zhang, Y.-T. Kao, W. Qiu, L. Wang, and D. Zhong, "Femtosecond Studies of Tryptophan Fluorescence Dynamics in Proteins: Local Solvation and Electronic Quenching," J. Phys. Chem. B 110(37), 18097–18103 (2006).

24. R. Jimenez, G.R. Fleming, P.V. Kumar, and M. Maroncelli, "Femtosecond solvation dynamics of water," Nature 369(6480), 471–473 (1994).

25. A. Barth, "Infrared spectroscopy of proteins," Biochimica et Biophysica Acta (BBA) - Bioenergetics 1767(9), 1073–1101 (2007).

26. M. Glasbeek, and H. Zhang, "Femtosecond Studies of Solvation and Intramolecular Configurational Dynamics of Fluorophores in Liquid Solution," Chem. Rev. 104(4), 1929–1954 (2004).

27. H. Chosrowjan, S. Taniguchi, N. Mataga, M. Unno, S. Yamauchi, N. Hamada, M. Kumauchi, and F. Tokunaga, "Low-Frequency Vibrations and Their Role in Ultrafast Photoisomerization Reaction Dynamics of Photoactive Yellow Protein," J. Phys. Chem. B 108(8), 2686–2698 (2004).

28. J. Xu, and J.R. Knutson, "Chapter 8 Ultrafast Fluorescence Spectroscopy via Upconversion," Methods in Enzymology, 159–183 (2008).

29. Z. Yu, L. Gundlach, and P. Piotrowiak, "Efficiency and temporal response of crystalline Kerr media in collinear optical Kerr gating," Opt. Lett. 36(15), 2904 (2011).

30. A.-H. Fattah, A.M. Flatae, A. Farrag, and M. Agio, "Ultrafast single-photon detection at high repetition rates based on optical Kerr gates under focusing," Opt. Lett. 46(3), 560 (2021).

31. N. Sugimoto, H. Kanbara, S. Fujiwara, K. Tanaka, Y. Shimizugawa, and K. Hirao, "Third-order optical nonlinearities and their ultrafast response in $Bi_2O_3$–$B_2O_3$–$SiO_2$ glasses," J. Opt. Soc. Am. B 16(11), 1904 (1999).

32. T. Lin, Q. Yang, J. Si, T. Chen, F. Chen, X. Wang, X. Hou, and K. Hirao, "Ultrafast nonlinear optical properties of $Bi2O3$–$B2O3$–$SiO2$ oxide glass," Optics Communications 275(1), 230–233 (2007).

33. N. Sugimoto, K. Hirao, H. Kanbara, S. Fujiwara, and K. Tanaka, "Ultrafast response of third-order optical nonlinearity in glasses containing $Bi_2O_3$," Opt. Lett. 21(20), 1637 (1996).

34. N. Sugimoto, S. Ito, S. Fujiwara, T. Suzuki, H. Kanbara, and K. Hirao, "Femtosecond and terahertz optical Kerr shutter switching in glass containing high concentration of $Bi2O3$," Optics Communications 161(1–3), 47–50 (1999).

35. W. Tan, H. Liu, J. Si, and X. Hou, "Control of the gated spectra with narrow bandwidth from a supercontinuum using ultrafast optical Kerr gate of bismuth glass," Applied Physics Letters 93(5), (2008).

36. L. Yan, J. Si, F. Chen, S. Jia, Y. Zhang, and X. Hou, "Pump power dependence of Kerr signals in femtosecond cross pump-probe optical Kerr measurements," Opt. Express 17(24), 21509 (2009).

37. T. Hasegawa, T. Nagashima, and N. Sugimoto, "Z-scan study of third-order optical nonlinearities in bismuth-based glasses," Optics Communications 250(4–6), 411–415 (2005).

38. J. Zhang, Q. Nie, S. Dai, T. Xu, F. Chen, X. Shen, and X. Wang, "Nonlinear optical properties in bismuth-based glasses," J. Wuhan Univ. Technol.-Mat. Sci. Edit. 26(1), 61–64 (2011).

39. J. Tauc, R. Grigorovici, and A. Vancu, "Optical Properties and Electronic Structure of Amorphous Germanium," Physica Status Solidi (b) 15(2), 627–637 (1966).

40. J. Tauc, "Optical properties and electronic structure of amorphous Ge and Si," Materials Research Bulletin 3(1), 37–46 (1968).

41. Y. Gao, J.-J. Ma, Y. Chen, and M.-H. Wang, "Effect of $Bi_2O_3$ on the structure and thermal properties of $Bi_2O_3$-$SiO_2$-$B_2O_3$ glasses prepared by sol-gel method," J Sol-Gel Sci Technol 103(3), 713–721 (2022).

42. Möncke, D., & Ehrt, D. (2021). Charge transfer transitions in glasses - Attempt of a systematic review. Optical Materials: X, 12, Article 100092.